\documentclass{article}
\usepackage{amssymb}
\usepackage{graphicx}
\usepackage{amsmath}

\newtheorem{theorem}{Theorem} [section]

\newtheorem{corollary}[theorem]{Corollary}

\newtheorem{example}[theorem]{Example}

\newtheorem{proposition}[theorem]{Proposition}
\newtheorem{remark}[theorem]{Remark}

\newenvironment{proof}[1][Proof]{\textbf{#1.} }{\ \rule{0.5em}{0.5em}}
\renewcommand{\bibitem}[2]{{\hspace*{-.75cm} #2}\smallskip \newline }
\allowdisplaybreaks[1]
\begin{document}

\author{Luis A. Guardiola\thanks{
Operations Research Center. Universidad Miguel Hern\'{a}ndez, Edificio
Torretamarit. Avda. de la Universidad s.n. 03202 Elche (Alicante), Spain.
E-mail: ana.meca@umh.es} \thanks{%
Corresponding author.}, Ana Meca$^{\dag }$ and Justo Puerto\thanks{%
Facultad de Matem\'{a}ticas, Universidad de Sevilla, 41012 Sevilla, SPAIN.
e-mail: puerto@us.es}}
\title{Production-inventory games and pmas games: characterizations of the
Owen point\thanks{
The research of the authors is partially supported by Spanish Ministry of
Education and Science, Junta de Andaluc{\'{\i}}a and Generalitat Valenciana
grants number: MTM2004-0909, HA2003-0121, MTM2005-09184-C02-02, ACOMP06/040,
CSD2006-00032, P06-FQM-01366. Authors acknowledge the useful comments made
by the Associate Editor and the referee.}}
\maketitle

\begin{abstract}
Production-inventory games were introduced in Guardiola et al. (2007) as a
new class of totally balanced combinatorial optimization games. From among
all core-allocations, the Owen point was proposed as a specifically
appealing solution. In this paper we study some relationships of the class
of production-inventory games and other classes of new and known games. In
addition, we propose three axiomatic characterizations of the Owen point. We
use eight axioms for these characterizations, among those, inessentiality
and additivity of players' demands are used for the first time in this paper.

\textbf{Key words:} production-inventory games -- Owen point -- totally
balanced combinatorial optimization games -- core-allocations

\textbf{2000 AMS Subject classification:} 91A12, 90B05
\end{abstract}

\section{Introduction}

In this paper we focus on the class of production-inventory cooperative
games, introduced in Guardiola et al. (2007). We consider a group of firms
that produce indivisible goods over a finite planning horizon to cover a
known demand. We assume that there exist three different types of costs:
production, inventory holding and backlogging. The goal of each individual
firm is to satisfy its entire demand over the planning horizon at a minimum
operation cost. (This model is known in the O.R. literature as the
Wagner-Whitin model, see Wagner and Whitin 1958).

Cooperation enters the model throughout coordination among firms.
Specifically, if a group of firms forms a coalition (joint venture) then
they will use the best technology among the members. This means that the
members of that coalition produce, hold inventory and pay backlogged demand
at the minimum cost of the coalition members. It is clear that the above
coordination process induces savings and therefore, studying the problem of
how to allocate the overall saving among the firms is a meaningful problem.
This allocation problem can be modeled by a transferable utility cooperative
game. In this game the worth of each coalition of firms is obtained solving
the combinatorial optimization problem that results from Wagner-Whitin model
with production, inventory holding and backlogging costs induced by the
members of the coalition. Other classes of combinatorial optimization games
can be found in Deng et al. (1999) and (2000) and the references therein.

The analysis of coordination in inventory problems is not new. Scanning the
literature, one can find centralization inventory models analyzed from this
point of view. Eppen (1979), Hartman et al. (2000), Hartman and Dror (2003)
and (2005), Slikker et al. (2005), and M\"{u}ller et al. (2002) have treated
cooperation in a news-vendor problem. A similar consideration for the
continuous review inventory model is studied in Gerchak and Gupta (1991),
Robinson (1993), and Hartman and Dror (1996). Tijs et al. (2005) studies a
situation where one agent owns an amount of storage space and the other
agents have some goods, part of which can be stored generating benefits. A
general framework for the study of continuous time decentralized
distribution systems is analyzed in Anupindi et al.(2001). The problem of
sharing the benefits produced by full cooperation between agents is tackled
by introducing a related cooperative game. Minner (2006) analyzes horizontal
cooperations between organizations that have the opportunity to jointly
replenish material requirements. In Meca et al. (2003), Meca et al. (2004)
and Meca (2007) a group of firms dealing with the ordering and holding of a
certain commodity (every individual agent's problem being an EPQ and EOQ
problem, respectively), either decide to cooperate and make their orders
jointly or consider coordination with regard to ordering and holding costs.
The interested reader is referred to Borm et al. (2001) for a detailed
presentation of inventory games, as well as other Operations Research games.


Production-inventory games (henceforth: PI-games) were studied in
Guar\-dio\-la et al. (2007) where it is shown that this class of games is
strictly included in the class of totally balanced games. It was also proven
that the Owen set (the set of allocations that are achievable through dual
solutions, see Owen 1975 and Gellekom et al. 2000) reduces to a singleton.
This fact motivates the name Owen point rather than Owen set within this
class of games. Finally, Guardiola et al. (2007) also proves that the Owen
point always belongs to the core of the game and that can be reached through
a population monotonic allocation scheme. Hence, every PI-game is a
non-negative cost game allowing for population monotonic allocation schemes
(henceforth: PMAS-game).

In this paper we prove that the class of PI-games coincides with the class
of PMAS-games, and we provide an interesting relationship between PI-games
and concave games. In addition, we present three different axiomatic
characterizations of the Owen point. To achieve the two first
characterizations we have kept in mind the work by Gellekom et al. (2000) in
which the Owen set of linear production games is characterized. The third
one, which is based on a population monotonicity property, is very natural
due to the fact that the class of PI-games coincides with the class of
non-negative cost games with a population monotonic allocation scheme.

The paper is organized as follows. We start by introducing definitions and
notation in Section \ref{preli}. In Section \ref{pmasg} we introduce the
class of PMAS-games. There, we also study relationships with PI-games and
other classes of games. Section \ref{axiomatic} provides three
characterizations of the Owen point. The paper finishes with a section
devoted to draw some conclusions and final remarks.

\section{\label{preli}Preliminaries}

A TU cost game is a pair $(N,c)$, where $N=\left\{ 1,2,...,n\right\} $ is
the finite set of players, $\mathcal{P}(N)$ is the set of nonempty
coalitions of $N$ and $c:\mathcal{P}(N)\rightarrow \mathbb{R}$ the
characteristic function satisfying $c(\varnothing )=0.$ The subgame related
to coalition $S,c_{S},$ is the restriction of the mapping $c$ to the
subcoalitions of $S.$ We denote by $\left\vert S\right\vert $ the cardinal
of set $S$, for all $S\subseteq N.$ A game $(N,c)$ is said to be 0-monotone
if it is monotone after 0-normalization, i.e., for all $S,S^{\prime }\subset
N$ with $S\subset S^{\prime }$ we have $c(S)-\sum_{i\in S}c(\{i\})\geq
c(S^{\prime })-\sum_{i\in S^{\prime }}c(\{i\}).$ A player $i\in N$ is a veto
player in the game $c$ if and only if $c(S)=0$ for all $S\subseteq
N\backslash \{i\}.$ For convenience we call a game with at least one veto
player a veto game. Finally, $(N,c)$ is called a simple game whenever $%
c(S)\in \{0,1\}$ for all $S\subseteq N.$

A cost-sharing vector will be $x\in \mathbb{R}^{N}$ and, for every coalition
$S\subseteq N$ we shall write $x_{S}:=\sum_{i\in S}x_{i}$, the cost-sharing
of coalition $S$ (where $x_{\varnothing }=0).$\textbf{\ }The core of the
game $(N,c)$ consists of those cost-sharing vectors which allocate the cost
of the grand coalition in such a way that every other coalition pays at most
its cost by the characteristic function: $Core(N,c)=\{x\in \mathbb{R}%
^{n}\left/ x_{N}=c(N)\text{ and }x_{S}\leq c(S)\text{ for all }S\subset
N\right. \}.$ In the following, cost-sharing vectors belonging to the core
will be called core-allocations. A cost game $(N,c)$ has a nonempty core if
and only if it is balanced (see Bondareva 1963 or Shapley 1967). It is a
totally balanced game (Shapley and Shubik, 1969) if the core of every
subgame is nonempty.

A well-known class of balanced games is the class of concave games (Shapley,
1971). A TU game $(N,c)$ is concave if and only if $c(S\cup \{i\})-c(S)\geq
c(T\cup \{i\})-c(T)\ $for all players $i\in N$ and all pairs of coalitions $%
S,T\subseteq N$ such that $S\subseteq T\subseteq N\backslash \{i\}.$ A
population monotonic allocation scheme (see Sprumont, 1990), or PMAS, for
the game $(N,c)$ is a collection of vectors $y^{S}\in \mathbb{R}^{\left\vert
S\right\vert }$ for all $S\subseteq N,S\neq \varnothing $ such that $%
y^{S}(S)=c(S)$ for all $S\subseteq N,S\neq \varnothing ,$ and $y_{i}^{S}\geq
y_{i}^{T}$ for all $S\subseteq T\subseteq N$ and $i\in S.$ Finally, for a
generic mathematical programming problem $(P),val(P)$ denotes the optimal
value of problem $(P)$.

A production-inventory situation (henceforth: PI-situation) is the one in
which several agents facing each one a production-inventory problem, decide
to cooperate to reduce costs. The cooperation is driven by sharing
technologies in production, inventory carrying and backlogged demand. Thus,
if a group of agents agree on cooperation then at each period they will
produce and pay inventory carrying and backlogged demand at the cheapest
costs among the members of the coalition. Formally, let $U$\ be an infinite
set, the universe of players. A PI-situation is a 3-tuple $(N,D,\Re )$ where
$N\subset U$ is a finite set of players $\left( \left\vert N\right\vert
=n\right) $, being $D$ an integer matrix of demands and $\Re =(H|B|P)$ is a
cost matrix, so that
\begin{equation*}
D=[d^{1},\ldots ,d^{n}]^{\prime },\quad H=[h^{1},\ldots ,h^{n}]^{\prime
},\quad B=[b^{1},\ldots ,b^{n}]^{\prime },\quad P=[p^{1},\ldots
,p^{n}]^{\prime };
\end{equation*}%
where:

\begin{itemize}
\item $T$ is the planning horizon.

\item $d^{i}=[d_{1}^{i},\ldots ,d_{T}^{i}]\geq 0,d_{t}^{i}=$ demand of the
player $i$ during period $t$, $t=1,\ldots ,T$.

\item $h^{i}=[h_{1}^{i},\ldots ,h_{T}^{i}]\geq 0,h_{t}^{i}=$ unit inventory
carrying costs of the player $i$ in period $t$, $t=1,\ldots ,T$.

\item $b^{i}=[b_{1}^{i},\ldots ,b_{T}^{i}]\geq 0,b_{t}^{i}=$ unit
backlogging carrying costs of the player $i$ in period $t$, $t=1,\ldots ,T$.

\item $p^{i}=[p_{1}^{i},\ldots ,p_{T}^{i}]\geq 0,p_{t}^{i}=$ unit production
costs of the player $i$ in period $t$, $t=1,\ldots ,T$.

The decision variables of the model, which are required to be integer
quantities, are:

\item $q_{t}=$ production during period $t$.

\item $I_{t}=$ inventory at hand at the end of period $t$.

\item $E_{t}=$ backlogged demand at the end of period $t$.
\end{itemize}

Note that we can associate with each PI-situation $(N,D,\Re )$ a cost
TU-game $(N,c)$ with characteristic function $c$ defined as follows: $%
c(\varnothing )=0$ and for any $S\subseteq N,c(S)=val(PI(S))$, where $\left(
PI(S)\right) $ is the following problem
\begin{eqnarray*}
(PI(S))\quad &\min
&\sum_{t=1}^{T}(p_{t}^{S}q_{t}+h_{t}^{S}I_{t}+b_{t}^{S}E_{t}) \\
&\mbox{s.t.}&I_{0}=I_{T}=E_{0}=E_{T}=0, \\
&&I_{t}-E_{t}=I_{t-1}-E_{t-1}+q_{t}-d_{t}^{S},\quad t=1,\ldots ,T, \\
&&q_{t},\;I_{t},\;E_{t},\text{ non-negative, integer, }t=1,\ldots ,T;
\end{eqnarray*}%
with
\begin{equation*}
p_{t}^{S}=\min_{i\in S}\{p_{t}^{i}\},\;h_{t}^{S}=\min_{i\in
S}\{h_{t}^{i}\},\;b_{t}^{S}=\min_{i\in
S}\{b_{t}^{i}\},\;d_{t}^{S}=\sum_{i\in S}d_{t}^{i}.
\end{equation*}

Every cost TU-game defined in this way is what we call a
production-inventory game. Guardiola et al. (2007) observes that the linear
relaxation, $\left( LPI(S)\right) $ of the problem $\left( PI(S)\right) ,$
has integer optimal solutions\ provided that the demands are integer. It
follows from standard duality theory of linear programming that the dual
problem for any coalition $S\subseteq N$ is the following problem,
\begin{eqnarray*}
(DLPI(S))\quad &\max &\sum_{t=1}^{T}d_{t}^{S}y_{t}\mbox{   } \\
&\mbox{s.t.}&y_{t}\leq p_{t}^{S},\qquad \qquad t=1,\ldots ,T, \\
&&y_{t+1}-y_{t}\leq h_{t}^{S},\quad t=1,\ldots ,T-1, \\
&&-y_{t+1}+y_{t}\leq b_{t}^{S},\quad t=1,\ldots ,T-1.
\end{eqnarray*}

In order to illustrate the structure of the model, we consider a special
case with two periods. Let $S$ be a coalition of players in $N$. The problem
$LPI(S)$ and its dual can be written as:
\begin{equation*}
\hspace*{-1cm}%
\begin{array}{rlcll}
\min & p_{1}^{S}q_{1}+p_{2}^{S}q_{2}+h_{1}^{S}I_{1}+b_{1}^{S}E_{1} &
\mbox{
 and } & \quad \max & d_{1}^{S}y_{1}+d_{2}^{S}y_{2} \\
\mbox{s.t.} & \left[
\begin{array}{rrrr}
1 & 0 & -1 & 1 \\
0 & 1 & 1 & -1%
\end{array}
\right] \left[
\begin{array}{c}
q_{1} \\
q_{2} \\
I_{1} \\
E_{1}%
\end{array}
\right] =\left[
\begin{array}{c}
d_{1}^{S} \\
d_{2}^{S}%
\end{array}
\right] & \quad & \quad \mbox{s.t.} & \left[
\begin{array}{rr}
1 & 0 \\
0 & 1 \\
-1 & 1 \\
1 & -1%
\end{array}
\right] \left[
\begin{array}{c}
y_{1} \\
y_{2}%
\end{array}
\right] \leq \left[
\begin{array}{c}
p_{1}^{S} \\
p_{2}^{S} \\
h_{1}^{S} \\
b_{1}^{S}%
\end{array}
\right] . \\
& q_{1},\;q_{2},\;I_{1},\;E_{1}\geq 0, & \quad &  &
\end{array}%
\end{equation*}

Guardiola et al. (2007)\ proves that the optimal solution of problem $%
(DLPI(S))$ is $y_{t}^{\ast }(S)=\min \Big\{p_{t}^{S},\min_{k<t}%
\{p_{k}^{S}+h_{kt}^{S}\},\min_{k>t}\{p_{k}^{S}+b_{tk}^{S}\}\Big\}$, for all $%
t=1,\ldots ,T,$ with
\begin{eqnarray*}
p_{k}^{S} &=&\left\{
\begin{array}{cc}
p_{1}^{S} & \text{if }k<1, \\
p_{T}^{S} & \text{if }k>T,%
\end{array}%
\right. \\
h_{kt}^{S} &=&\sum_{r=k}^{t-1}h_{r}^{S},\quad \mbox{for any
}k<t,t=2,\ldots ,T;h_{k1}^{S}=0,k<1, \\
b_{tk}^{S} &=&\sum_{r=t}^{k-1}b_{r}^{S},\quad \mbox{ for any }%
k>t,\;t=1,\ldots ,T-1;b_{Tk}^{S}=0,k>T.
\end{eqnarray*}

Moreover, these optimal solutions satisfy the following property: $%
y_{t}^{\ast }(S)\geq y_{t}^{\ast }(R)$ for all $S\subseteq R\subseteq N$\
and all $t\in \{1,...,T\}$. The characteristic function of PI-games can be
rewritten as follows: for any $\varnothing \neq S\subseteq
N,c(S)=\sum_{t=1}^{T}d_{t}^{S}y_{t}^{\ast }\left( S\right) $. \label{p:seis}

The reader may note that PI-games are not concave in general (see Example
4.4 in Guardiola et al. 2007). In these games the allocation $\left(
\sum_{t=1}^{T}d_{t}^{i}y_{t}^{\ast }\left( N\right) \right) _{i\in N}%
\hspace*{-0.3cm}=Dy^{\ast }(N)$ is called the Owen point, and it is denoted
by $Owen(N,D,\Re )$. At times, if there is no confusion, we simply use $o$
to refer to the Owen point. Following Guardiola et al. (2007), it turns out
that the Owen point is a core-allocation which can be reached through a
population monotonic allocation scheme; hence every PI-game is a totally
balanced game.

In some situations we will use $c^{(N,D,\Re )}(S)$ instead of $c(S)$, in
order to denote that the game $\left( N,c\right) $ comes from the situation $%
(N,D,\Re )$.

We denote by $\Upsilon $ the set of production-inventory situations $%
(N,D,\Re )$ defined in the universe of players $U$, being $n\geq 1,T\geq 1$
and $D$ an integer matrix. We say that a player $i\in N$ is essential if
there exists $t\in \{1,...,T\}$ with $d_{t}^{N\backslash \{i\}}>0$ such that
$y_{t}^{\ast }(N\backslash \{i\})>y_{t}^{\ast }(N)$. The reader may note
that an essential player is the one for which there exists at least one
period in which he is needed by the rest of players in order to produce at a
minimum cost a certain demand. We denote by $\mathcal{E}$ the set of
essential players. The players not being essential are called inessential.
It can be checked that for each inessential player $i$,\ $o_{N\setminus
\{i\}}=c(N\setminus \{i\})$ holds. Finally, Guardiola et al. (2007) showed
that the core of PI-games shrinks to the Owen point just only when all
players are inessential for the PI-situation.

\section{\label{pmasg}PI-games and PMAS-games}

This section introduces a new class of TU games closely related to PI-games,
namely PMAS-games. This class consists of all non-negative cost games
allowing for pmasses. It turns out that the classes of PI-and PMAS-games
coincide.

We start this section proving that the class of PI-games is closed under
finite sums. This result will be used later in the proof of the main theorem
of this section.

\begin{proposition}
\label{t:suma_PI} The sum of PI-games defined on the same set of players $N$
is a PI-game.
\end{proposition}

\begin{proof}
Consider two PI-games $\left( N,\overline{c}\right) $ and $\left( N,\widehat{%
c}\right) $ that arise from two PI-situations $(N,\overline{D},\overline{\Re
})$ and $(N,\widehat{D},\widehat{\Re })$, respectively. Denote by $\overline{%
T},\widehat{T}$ the number of periods for the first and second
PI-situations, respectively. Now, we build a new PI-situation $(N,D,\Re )$
with,

\begin{equation*}
D=\left( \overline{D}\left\vert 0\right\vert \widehat{D}\right) =\left(
\begin{array}{ccccccccc}
\overline{d}_{1}^{1} & \overline{d}_{2}^{1} & ... & \overline{d}_{\overline{T%
}}^{1} & 0 & \widehat{d}_{1}^{1} & \widehat{d}_{2}^{1} & ... & \widehat{d}_{%
\widehat{T}}^{1} \\
\overline{d}_{1}^{2} & \overline{d}_{2}^{2} & ... & \overline{d}_{\overline{T%
}}^{2} & 0 & \widehat{d}_{1}^{2} & \widehat{d}_{2}^{2} & ... & \widehat{d}_{%
\widehat{T}}^{2} \\
\vdots & \vdots & \vdots & \vdots & 0 & \vdots & \vdots & \vdots & \vdots \\
\overline{d}_{1}^{n} & \overline{d}_{2}^{n} & ... & \overline{d}_{\overline{T%
}}^{n} & 0 & \widehat{d}_{1}^{n} & \widehat{d}_{2}^{n} & ... & \widehat{d}_{%
\widehat{T}}^{n}%
\end{array}%
\right)
\end{equation*}

\begin{equation*}
H=\left( \overline{H}\left\vert \omega \right\vert \widehat{H}\right)
=\left(
\begin{array}{ccccccccc}
\overline{h}_{1}^{1} & \overline{h}_{2}^{1} & ... & \overline{h}_{\overline{T%
}}^{1} & \omega & \widehat{h}_{1}^{1} & \widehat{h}_{2}^{1} & ... & \widehat{%
h}_{\widehat{T}}^{1} \\
\overline{h}_{1}^{2} & \overline{h}_{2}^{2} & ... & \overline{h}_{\overline{T%
}}^{2} & \omega & \widehat{h}_{1}^{2} & \widehat{h}_{2}^{2} & ... & \widehat{%
h}_{\widehat{T}}^{2} \\
\vdots & \vdots & \vdots & \vdots & \omega & \vdots & \vdots & \vdots &
\vdots \\
\overline{h}_{1}^{n} & \overline{h}_{2}^{n} & ... & \overline{h}_{\overline{T%
}}^{n} & \omega & \widehat{h}_{1}^{n} & \widehat{h}_{2}^{n} & ... & \widehat{%
h}_{\widehat{T}}^{n}%
\end{array}%
\right)
\end{equation*}%
where $\omega \in \mathbb{R}$ is sufficiently large. Matrices $B$ and $P$
are defined in the same way as $H$. The PI-situation $(N,D,\Re )$ has $T=%
\overline{T}+\widehat{T}+1$ periods. Take a coalition $S\subseteq N$ and
consider $\overline{y}(S)\in R^{\overline{T}}$, $\widehat{y}(S)\in R^{%
\widehat{T}}$ being optimal solutions for the PI-situations $(N,\overline{D},%
\overline{\Re })$ and $(N,\widehat{D},\widehat{\Re })$, respectively. The
reader may note that is not optimal satisfying the demand $d_{t}^{S}$ with $%
t\leq \overline{T}$ in a period $\tilde{t}\geq \overline{T}+1$ since
backlogging costs are higher. By a similar argument the demand $d_{t}^{S}$
with $t>\overline{T}+1$ should not be satisfied from a period $\tilde{t}\leq
\overline{T}.$ Hence, the optimal solution for any coalition $S\subseteq N$
in a period $t\in \{1,...,T\}$ is given by
\begin{equation*}
y_{t}^{\ast }(S)=\left\{
\begin{array}{cl}
\overline{y}_{t}(S), & \text{if }t\leq \overline{T}, \\
&  \\
\omega , & \text{if }t=\overline{T}+1, \\
&  \\
\widehat{y}_{t}(S), & \text{if }\overline{T}+1<t\leq \widehat{T},%
\end{array}%
\right.
\end{equation*}%
being $\overline{y}_{t}(S)$ and $\widehat{y}_{t}(S)$ as defined above.
Hence, for each $S\subseteq N,$
\begin{equation*}
\overline{c}(S)+\widehat{c}(S)=\sum_{t=1}^{\overline{T}}\overline{d}_{t}^{S}%
\overline{y}_{t}(S)+\sum_{t=1}^{\widehat{T}}\widehat{d}_{t}^{S}\widehat{y}%
(S)=\sum_{t=1}^{T}d_{t}^{S}y_{t}^{\ast }(S)=c(S).
\end{equation*}%
\hfill
\end{proof}

Next we define the concept of PMAS-game. Every non-negative cost game $(N,c)$
that possesses a PMAS is called a PMAS-game. The reader may note that the
class of PMAS-games is a cone.

Our main result in this section states the relationship between PI-games and
PMAS-games.

\begin{theorem}
Every PMAS-game is a PI-game.
\end{theorem}

\begin{proof}
Take $(N,c)$ a PMAS-game. Due to Proposition \ref{t:suma_PI} we can assume
that $c$ is an extreme direction of the cone of PMAS-games. Sprumont (1990)
shows that (up to normalization) the set of extreme directions consists of
all 0-monotone simple veto games. Let $i_{v}$ be a veto player of $c$.
Define $M=\left\{ \left. S\subset N\right\vert c(S)=1\text{ and }c(S^{\prime
})=0\text{ for all }S^{\prime }\supset S\right\} ,$ i.e., $M$ consists of
all maximal coalitions with value 1. Let $T=\left\vert M\right\vert +1$ and
enumerate the elements of $M$, i.e., $M=\left\{ S_{1},...,S_{T-1}\right\} .$
Define $(N,D,\Re )$ by

\begin{equation*}
\begin{array}{cc}
d_{t}^{i}=\left\{
\begin{array}{cc}
1 & \text{if }(i,t)=(i_{v},T), \\
0 & \text{otherwise,}%
\end{array}%
\right. , & b_{t}^{i}=1\text{ for all }i\text{ and }t\text{,} \\
h_{t}^{i}=\left\{
\begin{array}{cc}
1 & \text{if }t<T\text{ and }i\in S_{t}, \\
0 & \text{otherwise,}%
\end{array}%
\right. , & p_{t}^{i}=\left\{
\begin{array}{cc}
0 & \text{if }t=1, \\
1 & \text{otherwise.}%
\end{array}%
\right. .%
\end{array}%
\end{equation*}

Let $(N,\overline{c})$ be the game generated by this PI-situation. We will
show that $\overline{c}$ and $c$ coincide. Let $S\subseteq N.$ If $%
i_{v}\notin S,$ then $\overline{c}(S)=c(S)=0,$ since $\sum_{j\in S}d^{j}=0$
and $i_{v}$ is a veto player of $c$. If $i_{v}\in S,$ then $S$ has to
produce one unit. If period $T$ is chosen to do this, it will cost 1. Hence,
$\overline{c}(S)\in \{0,1\}.$ Coalition $S$ can (only) produce freely in the
first period. It can fulfil the demand for free if it has no holding costs
in each period (but the last). This is the case if and only if $S\nsubseteq
S_{t}$ for every $t\in \{1,...,T-1\}.$ On the other hand, because $c$ is
0-monotone, $c(S)=1$ if and only if $S$ is a subset of some maximal
coalition with value $1$. We conclude that $\overline{c}$ and $c$
coincide.\medskip
\end{proof}

We conclude this section with a nice relationship between PI-games and
concave games.

\begin{corollary}
Each non-negative concave game is a PI-game.
\end{corollary}

\begin{proof}
It is follows directly from the fact that any non-negative concave game has
a PMAS: the Shapley value.
\end{proof}

\section{\label{axiomatic}Characterizations of the Owen point}

This section is devoted to address the second goal of this paper, namely to
provide axiomatic foundations for the Owen point. Axiomatic
characterizations of solution concepts in game theory is a fruitful area of
research. The interested reader is referred to Peleg and Sudh\"{o}lter
(2003) and Moulin (1988) for a comprehensive study of this field.

A solution rule $\varphi $ on $\Upsilon $ is a map, which assigns to every
production-inventory situation $(N,D,\Re )\in \Upsilon $ a subset of $%
\mathbb{R}^{n}$. In particular, the core and the Owen set are solution
rules. (Recall that the core was introduced by Gillies 1959 and the Owen set
by Owen 1975).

Gellekom et al. (2000) prove that the Owen set as considered as a solution
rule is not \textquotedblleft game-theoretical". Thus, first of all we
wonder whether the Owen point exhibits the same behavior.

Our next example shows that the Owen point is not a game-theoretical
solution since it depends on PI-situations and not on PI-games.

\begin{example}
\label{ex:no1} Consider the following PI-situation $(N,D,\Re )\in \Upsilon $
with two periods and two players, namely $P1$ and $P2$:

\begin{equation*}
\begin{tabular}{|c|c|c||c|c||c||c|}
\hline
& \multicolumn{2}{|c||}{Demand} & \multicolumn{2}{|c||}{Production} &
Inventory & Backlogging \\ \hline
P1 & 1 & 0 & 2 & 1 & 1 & 1 \\ \hline
P2 & 0 & 2 & 1 & 1 & 2 & 2 \\ \hline
\end{tabular}%
\end{equation*}

The data above gives rise to the game with characteristic function in the
following table:

\begin{equation*}
\begin{tabular}{|c|c|c||c|c||c||c||c|}
\hline
& $d_{1}^{S}$ & $d_{2}^{S}$ & $p_{1}^{S}$ & $p_{2}^{S}$ & $h_{1}^{S}$ & $%
b_{1}^{S}$ & $c$ \\ \hline
$\{1\}$ & 1 & 0 & 2 & 1 & 1 & 1 & 2 \\ \hline
$\{2\}$ & 0 & 2 & 1 & 1 & 2 & 2 & 2 \\ \hline
$\{1,2\}$ & 1 & 2 & 1 & 1 & 1 & 1 & 3 \\ \hline
\end{tabular}%
\end{equation*}

This gives $Owen(N,D,\Re )=\{(1,2)\}$. On the other hand, the PI-situation $%
(N,D^{\prime },\Re ^{\prime })\in \Upsilon $, where

\begin{equation*}
\begin{tabular}{|c|c|c||c|c||c||c|}
\hline
& \multicolumn{2}{|c||}{Demand} & \multicolumn{2}{|c||}{Production} &
Inventory & Backlogging \\ \hline
P1 & 0 & 2 & 1 & 1 & 2 & 2 \\ \hline
P2 & 1 & 0 & 2 & 1 & 3 & 2 \\ \hline
\end{tabular}%
\end{equation*}%
provides the same PI-game, but now the Owen point is different $%
Owen(N,D^{\prime },\Re ^{\prime })=\{(2,1)\}.$
\end{example}

In spite of the behavior shown by Example \ref{ex:no1}, the Owen point is an
attractive cost sharing vector in PI-games. Actually, it is the unique
core-allocation reached by dual solutions within the class of PI-games.
Moreover, there is always a PMAS that realizes the Owen point. Therefore,
our goal in the rest of the section is to find different axiomatic
characterizations for the Owen point.

Let $\varphi $ be a solution rule on $\Upsilon $, we consider the following
properties:

\begin{itemize}
\item[(EF)] \textit{Efficiency}. For all $y\in \varphi (N,D,\Re )$ and for
all $(N,D,\Re )\in \Upsilon $, $y_{N}=c^{(N,D,\Re )}(N)$.

\item[(NE)] \textit{Nonemptiness}. For all $(N,D,\Re )\in \Upsilon $, $%
\varphi (N,D,\Re )\neq \varnothing $.

\item[(PO)] \textit{Positivity.} For all $(N,D,\Re )\in \Upsilon $\ and for
all $y\in \varphi (N,D,\Re ),$ $y_{i}\geq 0$ for each $i\in N$.

\item[(IR)] \textit{Individual rationality}. For all $(N,D,\Re )\in \Upsilon
$, for all $y\in \varphi $ $(N,D,\Re )$, and for all $i\in N$, $y_{i}\leq
c^{(N,D,\Re )}(\{i\})$.

\item[(IE)] \textit{Inessentiality}. For all $(N,D,\Re )\in \Upsilon $ and
for all $y\in \varphi (N,D,\Re )$, if $i$ is an inessential player for $%
(N,D,\Re )$, then $y_{N\setminus \{i\}}\leq c^{(N,D,\Re )}(N\setminus \{i\})$%
.

\item[(AP)] \textit{Additivity of players' demands}. For all $(N,D,\Re )\in
\Upsilon $ and for all $y\in \varphi (N,D,\Re )$, there exists $%
(z_{k})_{k\in N}\in (\mathbb{R}^{n})^{n}$ such that $y=\sum_{k\in N}z_{k}$
and for all $k\in N$, $z_{k}\in \varphi (N,D_{k},\Re )$, where
\begin{equation}
D_{k}=\left( d^{it}\right) _{\substack{ i=1,...,n  \\ t=1,...,T}},\text{ }%
d^{it}=\left\{
\begin{array}{cc}
d_{t}^{i} & \text{if }k=i, \\
0 & \text{otherwise.}%
\end{array}%
\right.  \label{eq3}
\end{equation}
\end{itemize}

\textit{Efficiency} guarantees that the overall cost of any PI-situation is
entirely divided among the players. \textit{Nonemptiness} assures that the
solution rule will never return the empty set as an admissible allocation.
\textit{Positivity} ensures that the allocation process of the overall
saving does not compensate those players that have zero demand for all
periods (subsidizing players is not allowed). On the contrary, it forces the
remaining players to pay their own demands. \textit{Individual rationality}
protects all players from supporting a higher cost than what they generate
by themselves. \textit{Inessentiality} imposes collective rationality for
every coalition in which an inessential player has left.

Finally, a solution rule satisfies \textit{additivity of players' demands}
property if it is additive for the demand of each individual player; i.e. it
is additive with respect to matrices where just a player keeps its demand
and the remaining demands are set to zero. Note that the this particular
form of additivity arises from the fact that $c^{(N,D,\Re )}=\sum_{i\in
N}c^{(N,D_{i},\Re )}$ for all $(N,D,\Re )\in \Upsilon .$ Therefore, we focus
on those solution rules for PI-situations that are consistent with this kind
of partition of demands.

In order to get some insights into the above properties, the reader can
easily check that the Shapley value satisfies EF, NE and AP. In the
following, we also prove that the Owen point satisfies all the above
properties.

\begin{proposition}
\label{con}On $\Upsilon ,$ the Owen point satisfies EF, NE, PO, IR, IE, and
AP.
\end{proposition}

\begin{proof}
For all $(N,D,\Re )\in \Upsilon$ we have that $Owen(N,D,\Re )\in Core(N,c)$.
Then the Owen point satisfies EF, IR, IE and NE. Moreover, since all costs
and demands are non-negatives, $Owen_{i}(N,D,\Re
)=\sum_{t=1}^{T}d_{t}^{i}y_{t}^{\ast }(N)\geq 0$ for each $i\in N$.
Therefore, the Owen point satisfies PO.

Finally, if we define $D_{i}$ as in (\ref{eq3}), we have that $\sum_{i\in
N}D_{i}=D$ and
\begin{eqnarray*}
Owen(N,D,\Re ) &=&Dy^{\ast }(N)=\left( \sum_{i\in N}D_{i}\right) y^{\ast }(N)
\\
&=&\sum_{i\in N}\left( D_{i}y^{\ast }(N)\right) =\sum_{i\in
N}Owen(N,D_{i},\Re ).
\end{eqnarray*}%
Hence, the Owen point satisfies AP.
\end{proof}

\medskip

Now, we can prove that if all players are inessential, the Owen point can be
characterized just by three of the above properties.


\begin{proposition}
\label{con2}Let $(N,D,\Re )\in \Upsilon $ such that $\left\vert \mathcal{E}%
\right\vert =0$. The solution rule $\varphi $ on $\Upsilon $ satisfies EF,
NE, and IE if and only if $\varphi (N,D,\Re )=Owen(N,D,\Re )$.
\end{proposition}

\begin{proof}
(If) Immediately follows by Proposition \ref{con}. \medskip

\noindent (Only if) By NE, $\varphi (N,D,\Re )\neq \varnothing $ and by EF, $%
y_{N}=c^{(N,D,\Re )}(N)$ for all $y\in \varphi (N,D,\Re )$.

Take $y\in \varphi (N,D,\Re ).$ Since all players $i\in N$ are inessential,
by IE, it holds that $y_{N\setminus \{i\}}\leq c^{(N,D,\Re )}(N\setminus
\{i\})=o_{N\setminus \{i\}}$ for each $i\in N$. Hence, $y_{i}\geq
c^{(N,D,\Re )}(N)-c^{(N,D,\Re )}(N\setminus \{i\})=o_{N}-o_{N\setminus
\{i\}}=o_{i}$ for all $i\in N.$ Therefore by EF, $\varphi (N,D,\Re
)=Owen(N,D,\Re )$.\medskip
\end{proof}

The next two results of this section state that, in general, characterizing
the Owen point for a number of players $n,$ (arbitrary but fixed) can be
done with different combinations of the above six properties.

\begin{theorem}
\label{charac}Let $(N,D,\Re )\in \Upsilon .$ An allocation rule on $(N,D,\Re
)$ satisfies EF, NE, PO, IE, and AP if and only if it coincides with the
Owen point.
\end{theorem}

\begin{proof}
(If) Follows from Proposition \ref{con}. \medskip

\noindent (Only if) Let $\varphi $ be an allocation rule$.$ If $\left\vert
\mathcal{E}\right\vert =0$ by Proposition \ref{con2} $\varphi (N,D,\Re
)=Owen(N,D,\Re )$. Therefore, we suppose that $\left\vert \mathcal{E}%
\right\vert \geq 1$. We have that $D=D_{1}+D_{2}+...+D_{n}$ where $D_{i}$ is
(see (\ref{eq3})):
\begin{equation*}
D_{i}=\left(
\begin{array}{cccc}
0 & 0 & \dots & 0 \\
\vdots & \vdots & \vdots & \vdots \\
0 & 0 & \dots & 0 \\
d_{1}^{i} & d_{2}^{i} & \dots & d_{T}^{i} \\
0 & 0 & \dots & 0 \\
\vdots & \vdots & \vdots & \vdots \\
0 & 0 & \dots & 0%
\end{array}%
\right) .
\end{equation*}%
Then for all $i\in N$, $(N,D_{i},\Re )$ is a PI-situation with $D_{i}$ an
integer matrix and therefore it belongs to $\Upsilon $. Take $i\in N,$ then
the Owen point for $(N,D_{i},\Re )$ is given by $(o_{k})_{k=1,\ldots ,n}$:
\begin{equation}
o_{k}=\left\{
\begin{array}{cc}
\sum_{t=1}^{T}d_{t}^{i}y_{t}^{\ast }(N) & \text{if }k=i, \\
0 & \text{otherwise.}%
\end{array}%
\right.  \label{Owen}
\end{equation}

By NE, $\varphi (N,D_{i},\Re )\neq \varnothing $. Take $y\in \varphi
(N,D_{i},\Re )$, the player $i$\ is inessential for this situation, since
there is no $t\in \{1,...,T\}$\ with $d_{t}^{N\backslash \{i\}}>0$ such that
$y_{t}^{\ast }(N\backslash \{i\})<y_{t}^{\ast }(N).$\ Moreover, $%
c^{(N,D_{i},\Re )}(N\setminus \{i\})=0.$ By IE $y_{N\setminus \{i\}}\leq
c^{(N,D_{i},\Re )}(N\setminus \{i\})=0.$ Therefore, by PO, $y_{j}=0$\ for
each $j\in N\setminus \{i\}.$ Finally, by EF, $y=o$ and hence $\varphi
(N,D_{i},\Re )=Owen(N,D_{i},\Re )$. Thus, if $y\in \varphi (N,D,\Re )$ by AP
it follows that $y=z_{1}+...+z_{n}$ with $z_{i}\in \varphi (N,D_{i},\Re )$
for all $i\in N$, and so
\begin{equation*}
y=\sum_{i\in N}z_{i}=\sum_{i\in N}Owen(N,D_{i},\Re )=Owen(N,D,\Re ).
\end{equation*}

Hence, we conclude that $\varphi (N,D,\Re )=Owen(N,D,\Re )$.
\end{proof}

\begin{remark}
The reader may note that Theorem \ref{charac} could be proved without using
Proposition \ref{con2}. However, we propose this alternative proof in order
to stress that for those PI-situations where all players are inessential,
the Owen point (and hence the core) can be characterized by just three
properties.
\end{remark}

\begin{remark}
\label{remark}An alternative characterization for the Owen point can be
obtained just swapping properties IE and IR. It is clear that the Owen point
satisfies EF, NE, PO, IR, and AP. Now, let $\varphi $ be an allocation rule.
By NE, an allocation $y\in \varphi (N,D_{i},\Re )$ exists. Note that $%
c^{(N,D_{i},\Re )}(\{j\})=0$ for all $j\in N\setminus \{i\}$ then by PO and
IR, $y_{j}=0$ for each $j\in N\setminus \{i\}.$ Finally, by EF, $%
y_{i}=c^{(N,D_{i},\Re )}(N)=\sum_{t=1}^{T}d_{t}^{i}y_{t}^{\ast }(N)$ and
hence $\varphi (N,D_{i},\Re )=Owen(N,D_{i},\Re )$. Since both $\varphi $ and
the Owen point satisfy AP it follows that $\varphi (N,D,\Re )=Owen(N,D,\Re )$%
.
\end{remark}

In the following, we prove that the two sets of axioms used in Theorem \ref%
{charac} and Remark \ref{remark} are logically independent.

\begin{example}
Consider $\varphi $ on $\Upsilon $ defined by
\begin{equation*}
\varphi (N,D,\Re ):=\left\{
\begin{array}{cc}
\left\{
\begin{array}{c}
\left( \frac{c^{(N,D,\Re )}(N)}{2},\frac{c^{(N,D,\Re )}(N)}{2}\right) , \\
Owen(N,D,\Re )%
\end{array}%
\right\} , & (N,D,\Re )\in \Upsilon ^{1} \\
&  \\
Owen(N,D,\Re ), & \text{otherwise,}%
\end{array}%
\right.
\end{equation*}%
\noindent where
\begin{equation*}
\Upsilon ^{1}:=\left\{ (N,D,\Re )\in \Upsilon \left/
\begin{array}{c}
|N|=T=2, \\
c^{(N,D,\Re )}(N)\leq c^{(N,D,\Re )}(\{i\})\quad \forall i\in N%
\end{array}%
\right. \right\} .
\end{equation*}%
\noindent $\varphi (N,D,\Re )$ satisfies NE, EF, PO, IR and IE but not AP.
\end{example}

\begin{example}
We take $\varphi $ on $\Upsilon $ defined by
\begin{equation*}
\varphi (N,D,\Re ):=\left\{ \left( c^{(N,D,\Re )}(N),0,0,...,0\right) \in
\mathbb{R}^{N}\right\} .
\end{equation*}

\noindent {$\varphi $}$(N,D,\Re )${\ } satisfies NE, EF, PO and AP but
neither IR nor IE.
\end{example}

\begin{example}
Take $\varphi $ on $\Upsilon $ given by

\begin{equation*}
\varphi (N,D,\Re ):=\left\{
\begin{array}{cc}
\left( p_{1}^{1}\text{ }d_{1}^{1}+(p_{1}^{1}-p_{1}^{2})d_{1}^{2},p_{1}^{2}%
\text{ }d_{1}^{2}\right) , & (N,D,\Re )\in \Upsilon ^{2} \\
&  \\
Owen(N,D,\Re ), & \text{otherwise,}%
\end{array}%
\right.
\end{equation*}%
\noindent where
\begin{equation*}
\Upsilon ^{2}:=\left\{ (N,D,\Re )\in \Upsilon \left/ \left\vert N\right\vert
=2,T=1,p_{1}^{1}\text{ }<p_{1}^{2}\text{ }\right. \right\} .
\end{equation*}

\noindent {$\varphi $}$(N,D,\Re )${\ } satisfies NE, EF, IR, IE and AP but
not PO.
\end{example}

\begin{example}
Let $\varphi $ on $\Upsilon $ be defined by
\begin{equation*}
\varphi (N,D,\Re ):=\left\{ (0,0,...,0)\in \mathbb{R}^{N}\right\} ,
\end{equation*}

\noindent {$\varphi $}$(N,D,\Re )${\ } satisfies NE, PO, IR, IE and AP but
not EF.
\end{example}

\begin{example}
Define $\varphi $ on $\Upsilon $ by
\begin{equation*}
\varphi (N,D,\Re ):=\varnothing .
\end{equation*}

\noindent {$\varphi $}$(N,D,\Re )${\ satisfies EF, PO, IR, IE and AP but not
NE.}
\end{example}

To conclude this section we provide an alternative characterization for the
Owen point, based on a different rationale: consistency. In the previous
section we proved that the class of PI-games coincides with the class of
non-negative cost games with pmasses. Here, we will use a consistency
property on population monotonicity (See Thomson 1995 for further details on
population monotonic solution rules) that allows us to characterize the Owen
point. This characterization is similar to the one given in Grafe et al.
(1998) when studying the proportional rule on the class of externality games
(see Theorem 4.2) and is based on the following properties.

\begin{itemize}
\item[(PM)] Population monotonicity. A solution rule $\varphi $ defined on $%
\Upsilon $ is said to satisfy PM if for all $(N,D,\Re )\in \Upsilon $, for
all $S\subseteq N$, for all $y\in \varphi (N,D,\Re )$, $z\in \varphi
(S,D_{S},\Re _{S})$, and for all $i\in S$, we have $y_{i}\leq z_{i}$, being $%
(S,D_{S},\Re _{S})\in \Upsilon $ derived from $(N,D,\Re )$ by restricting
each matrix to the members of $S.$

\item[(AN)] Anonymity property. A solution rule $\varphi $ defined on $%
\Upsilon $ is said to satisfy AN if for every PI-situations $(N,D,\Re ),\;$%
every bijection $\sigma :N\longrightarrow N^{\prime },$ and every $y\in
\varphi (N,D,\Re ),$ we have $z\in (N^{\prime },D^{\prime },\Re ^{\prime }),$
where $y_{i}=z_{\sigma (i)}$ and $\sigma (i)$ has the demands and costs in $%
(N^{\prime },D^{\prime },\Re ^{\prime })$ that $i$ has in $(N,D,\Re ).(i\in
N)$
\end{itemize}

The last result of this section proves that there is a unique nonempty,
efficient and anonymous solution rule on the set of production-inventory
situations that satisfies population monotonicity: the Owen point.

\begin{theorem}
An allocation rule on $\Upsilon $ satisfies EF, NE, PM, and AN if and only
if it is the Owen point.
\end{theorem}

\begin{proof}
(If) It is clear that the Owen point satisfies EF, NE, AN and PM.

\noindent (Only if) Let $\varphi $ be an allocation rule on $\Upsilon $ that
satisfies EF, NE, PM, and AN. We assume that $\varphi \neq Owen.$ Then there
exists a PI-situation $(N,D,\Re )$ such that for some $i\in N$ and \textbf{%
for some} $y\in \varphi (N,D,\Re ),$ $y_{i}\neq Owen_{i}(N,D,\Re ).$ By EF%
\textbf{, }one has that $n\geq 2$. Besides, again by EF, there has to be a
player $j\in N$ \textbf{such that} $y_{j}<Owen_{j}(N,D,\Re ).$

Let $k$ be a player not in $N$. Consider a PI-situation $(\widetilde{N},%
\widetilde{D},\widetilde{\Re })$ where $\widetilde{N}=N\cup \{k\}$ and for
each $i\in \widetilde{N}\backslash \{k\},\widetilde{D}_{i}=D_{i},\widetilde{%
\Re }_{i}=\Re _{i}$ and $\widetilde{D}_{k}=D_{j},\widetilde{\Re }_{k}=\Re
_{j}.$ Then $y_{t}^{\ast }(N)=\widetilde{y}_{t}^{\ast }(\widetilde{N})$ for
all $t\in T.$ \textbf{Take} $z\in \varphi (\widetilde{N},\widetilde{D},%
\widetilde{\Re }).$ In addition, by PM, $z_{i}\leq y_{i}$, for each $i\in N.$
Moreover, by AN, $z_{j}=z_{k},$ so $z_{k}<Owen_{j}(N,D,\Re ).$ Then by EF we
obtain:%
\begin{equation*}
\sum_{t=1}^{T}\widetilde{d}_{t}^{\widetilde{N}}y_{t}^{\ast
}(N)=\sum_{t=1}^{T}\widetilde{d}_{t}^{\widetilde{N}}\widetilde{y}^{\ast }(%
\widetilde{N})=\sum_{i\in \widetilde{N}}z_{i}.
\end{equation*}%
Furthermore by EF and PM we obtain:
\begin{eqnarray*}
\sum_{i\in \widetilde{N}}z_{i} &=&z_{k}+z_{j}+\sum_{i\in N\backslash
\{j\}}z_{i}\leq y_{j}+y_{j}+\sum_{i\in N\backslash \{j\}}y_{i} \\
&<&Owen_{j}(N,D,\Re )+\sum_{i\in N}Owen_{i}(N,D,\Re )=\sum_{t=1}^{T}%
\widetilde{d}_{t}^{\widetilde{N}}y_{t}^{\ast }(N).
\end{eqnarray*}

Consequently there is a contradiction.
\end{proof}

\section{Concluding Remarks}

In this paper we have revisited the class of production-inventory games,
focussing on two of the open problems proposed in Guardiola et al. (2007):
1) Studying relationships of the class of PI-games with other classes of
games and 2) Finding axiomatic characterizations of the Owen point. First of
all, we have introduced the class of non-negative games with pmasses to
prove that PI-games and PMAS-games coincides. Regarding the second point, we
have introduced a set of axioms: \textit{efficiency}, \textit{nonemptiness},
\textit{positivity}, \textit{inessentiality}, \textit{individual rationality}%
, \textit{additivity of players' demands}, a\textit{nonymity}, and \textit{%
population monotonicity} that provide three different characterizations of
the Owen point for production-inventory situations.



\begin{thebibliography}{99}
\bibitem{Ze91} {Anupindi R, Bassok Y, Zemel E (2001) A general framework for
the study of decentralized distribution systems.\ Manufacturing \& Service
Operations Management 3:349-368}

\bibitem{B63} {\ Bondareva ON (1963) Some applications of linear programming
methods to the theory of cooperative games.\ Problemy Kibernety 10:119-139}

\bibitem{BHH01} {\ Borm PEM, Hamers H, Hendrickx R (2001) Operations
research games: a survey.\ TOP 9:139-216}

\bibitem{DIN99} {\ Deng X, Ibaraki T, Nagamochi H (1999) Algorithmic aspect
of the core of combinatorial optimization games.\ Math Oper Res 24:751-766}

\bibitem{DINZ00} {\ Deng X, Ibaraki T, Nagamochi H, Zang W (2000) Totally
balanced combinatorial optimization games.\ Math Programming 87:441-452}

\bibitem{E79} {\ Eppen GD (1979) Effect of centralization on expected cost
in a multi-location newsboy problem.\ Manage Sci 25:498-501}

%

\bibitem{GPRE00} {\ Gellekom JRG, Potters JAM, Reijnierse JH, Engel MC, Tijs
SH (2000) Characterization of the Owen set of Linear production processes.\
Games Econ Behav 32:139-156}

\bibitem{GG91} {\ Gerchak Y, Gupta D (1991) On apportioning costs to
customers in centralized continuous review inventory systems.\ J Oper Manage
10:546-551}

\bibitem{G59} {\ Gilles DB (1959) Solutions to general non-zero-sum games.\
Contributions to the Theory of Games vol. IV, Annals of Math Studies Vol.
40:47-85}


\bibitem{GIZ98} {\ Grafe F, I\~{n}arra E, Zarzuelo JM (1998) Population
monotonic allocation schemes on externality games. Math Methods Oper Res
48:71-80}

\bibitem{GMJ04} {\ Guardiola LA, Meca A, Puerto J (2007)
Production-Inventory games: a new class of totally balanced combinatorial
optimization games.\ Games Econ Behav (to appear)}


\bibitem{HM03} {\ Hartman BC, Dror M (2003)\ Optimizing centralized
inventory operations in a cooperative game theory setting.\ {\ IIE Trans
Oper Engineering} 35, 243-257}

\bibitem{HM05} {\ Hartman BC, Dror M (2005)\ Allocation of gains from
inventory centralization in newsvendor environments.\ {\ IIE Trans
Scheduling Logist} 37, 93-107}

\bibitem{HD96} {\ Hartman BC, Dror M (1996) Cost allocation in continuous
review inventory models.\ Naval Res Logist 43:549-561}

\bibitem{HDS00} {\ Hartman BC, Dror M, Shaked M (2000) Cores of inventory
centralization games. Games Econ Behav 31:26-49}

\bibitem{M04} {\ Meca A (2007) A core-allocation family for generalized
holding cost games.\ Math Methods Oper Res (to appear)}

\bibitem{MTGB04} {\ Meca A, Timmer J, Garc\'{\i}a-Jurado I, Borm PEM (2004)
Inventory games.\ European J Oper Res 156:127-139}

\bibitem{MGB03} {\ Meca A, Garc\'{\i}a-Jurado I, Borm PEM (2003) Cooperation
and competition in Inventory Games.\ Math Methods Oper Res 57:481-493}

\bibitem{M03} {\ Minner S (2006) Bargaining for cooperative economic
ordering. Decis Support Syst 43:569-583.}

\bibitem{MO88} {\ Moulin H (1988) Axioms of Cooperative Decision Making.\
Cambridge U. Press, New York.}

\bibitem{MSS02} {\ M\"{u}ller A, Scarsini M, Shaked M (2002) The newsvendor
game has a nonempty core.\ Games Econ Behav 38:118-126}

\bibitem{O75} {\ Owen G (1975) On the core of linear production games.\ Math
Programming 9:358-370}

\bibitem{PS03} {\ Peleg B, Sudh\"{o}ter P (2003) Introduction to the Theory
of Cooperative Games.\ Kluwer Academic, Boston.}

\bibitem{R93} {\ Robinson LW (1993). Comment on ``On apportioning costs to
customers in centralized continuous review inventory systems,'' \ by Gerchak
and Gupta. J Oper Manage 11:99-102}

\bibitem{SH67} {\ Shapley LS (1967)\ On balanced sets and cores.\ Naval Res
Logist 14:453-460}

\bibitem{SH71} {\ Shapley LS (1971)\ Cores of convex games. \ Int J Game
Theory 1:11-26}

\bibitem{SS69} {\ Shapley LS, Shubik M (1969)\ On market games.\ J Econ
Theory 1:9-25}

\bibitem{SFW01} {\ Slikker M, Fransoo J, Wouters M (2005)\ Cooperation
between multiple news-vendors with transshipments.\ European J Oper Res
167:370-380}

\bibitem{S90} {\ Sprumont Y (1990) Population monotonic allocation schemes
for cooperative games with transferable utility.\ Games Econ Behav 2:378-394}

\bibitem{T95} {\ Thomson W (1995) Population-monotonic allocation rules. \
Chapter 4 in \emph{Social Choice, Welfare and Ethics} (W.Barnett, H. Moulin,
M. Salles and N. Chofield, eds) Cambridge University Press, 79-124}

\bibitem{TML00} {\ Tijs SH, Meca A, L\'{o}pez MA (2005) Benefit sharing in
holding situations.\ European J Oper Res 162:251-269}

\bibitem{WW58} {\ Wagner HM, Whitin TM (1958) Dynamic version of the
economic lot size model.\ Manage Sci 5:89-96}
\end{thebibliography}
\end{document}